\documentclass{ws-ijmpe}

\newcommand{\ket}[1]{\ensuremath{\,|{#1}\rangle}}

\newcommand{\op}[1]{\ensuremath{\mathrm{#1}}}
\newcommand{\adj}[1]{\ensuremath{{{#1}}^{\dag}}}
\newcommand{\corr}[1]{\ensuremath{\widetilde{#1}}}

\newcommand{\HO}{\ensuremath{\op{H}}}

\newcommand{\TO}{\ensuremath{\op{T}}}
\newcommand{\VO}{\ensuremath{\op{V}}}

\newcommand{\PsiO}{\ensuremath{\op{\Psi}}}

\newcommand{\UCOM}{\ensuremath{\textrm{UCOM}}}

\begin{document}

\markboth{N. Paar, P. Papakonstantinou, R. Roth, and H. Hergert}{Self-consistent description of collective excitations in the unitary
correlation operator model}

\catchline{}{}{}{}{}

\title{SELF-CONSISTENT DESCRIPTION OF COLLECTIVE
EXCITATIONS IN THE UNITARY CORRELATION OPERATOR METHOD}
\author{N. PAAR\footnote{on leave of absence from Physics Department, Faculty of Science,
University of Zagreb, Croatia} \footnote{corresponding author: nils.paar@physik.tu-darmstadt.de}, P. PAPAKONSTANTINOU, R. ROTH, AND H. HERGERT}
\address{Institut f\"ur Kernphysik, Technische Universit\"at
  Darmstadt, \\
  Schlossgartenstrasse 9, D-64289 Darmstadt, Germany}

\maketitle

\begin{history}
\received{(received date)}
\revised{(revised date)}
\end{history}

\begin{abstract}
The fully self-consistent Random Phase Approximation (RPA) is constructed within
the Unitary Correlation Operator Method (UCOM), which
describes the dominant interaction-induced short-range central and tensor
correlations by a unitary transformation. Based on the correlated Argonne V18 interaction,
the RPA is employed in studies of multipole response in closed-shell nuclei across the nuclide chart.
The UCOM-RPA results in a collective character of giant resonances,
and it describes rather well the properties of isoscalar giant monopole resonances.
However, the excitation energies of
isovector giant dipole resonances and isoscalar giant quadrupole resonances
are overestimated due to the missing long-range correlations and
three-body contributions.
\end{abstract}

\section{Introduction}
One of the unresolved problems in the theory of nuclear structure is the description
of ground state properties and excitation phenomena for heavier nuclei,
based on realistic nucleon-nucleon (NN) interactions which reproduce the NN scattering 
data~\cite{Sto.93,Mac.89,Wir.95}. The use of these realistic
interactions for solving the nuclear many-body problem is a challenging task. Presently, 
only light nuclei can be treated within ab initio schemes like 
Green's function Monte Carlo \cite{Pie.04}, no-core shell model~\cite{Nav.03},
and  coupled cluster method~\cite{Wlo.05}. 
The Unitary Correlation Operator Method (UCOM), which describes the
dominant short-range and tensor correlations explicitly by means of a unitary transformation~\cite{Fel.98,Nef.03,Rot.04,Rot.05}, allows for the use of realistic NN interactions in traditional nuclear structure
methods. In contrast to other methods using unitary transformations, e.g. the unitary model operator
approach~\cite{Sha.67,Fuj.04}, the correlation operators are given explicitly
allowing for the derivation of a system-independent effective interaction operator $V_{\UCOM}$. 
Although different by its construction, the correlated NN interaction $V_{\UCOM}$ is similar to the $V_{low-k}$
low-momentum interaction \cite{Bog.03}. Both approaches lead to the separation 
of momentum scales, providing a phase-shift equivalent NN interaction in the low-momentum regime.

Very recently, the correlated realistic NN interaction constructed within 
the UCOM framework, has been employed in Hartree-Fock (HF) calculations across
the nuclide chart~\cite{RPPH.05}. Based on the UCOM-HF ground state, 
we have constructed the random-phase approximation (RPA) for the description of
low-amplitude collective excitations in atomic nuclei 
using correlated realistic NN interactions~\cite{PPRH.05}.
Various phenomenological RPA and quasiparticle RPA models have been very successful
in the past, not only in studies of giant resonances
and low-lying states (e.g. Refs.~\cite{Row.70,Dum.83,Daw.90,Ham.97,Col.00}),
but also in description of exotic nuclear structure of collective excitations
in nuclei away from the valley 
of $\beta$-stability~\cite{Ham.97,Mat.05,Ter.04,Paa.03,Sar.04,Paa_pp.05,Pap.04,Cao.05}.
The present study, however, provides the first insight into the collective
excitation phenomena in closed-shell nuclear systems,
based on the correlated realistic NN interactions.
\section{Unitary Correlation Operator Method (UCOM)}
The essential ingredient of the UCOM approach is the explicit treatment of the interaction-induced
correlations, i.e. short-range central and tensor correlations~\cite{Fel.98,Nef.03,Rot.04}.
The relevant correlations are imprinted into an uncorrelated many-body state $\ket{\Psi}$
through a state-independent unitary transformation defined by the
unitary operator $C$, resulting in a correlated state $ \ket{\corr{\PsiO}} = C \ket{\Psi} \;$.
An equivalent, technically more advantageous approach, is based on using the correlated 
operators $\widetilde{O}=\adj{C}OC$ with uncorrelated many-body states. 
The short-range central correlations are 
described by a distance-dependent shift, pushing two nucleons
apart from each other if they are closer than
the core distance.
The application of the correlation operator in two-body space
corresponds to a norm conserving coordinate transformation
with respect to the relative coordinate. This transformation is
parameterized in terms of correlation functions for each $(S,T)$ channel which are determined 
by an energy minimization in the two-body system. For purely repulsive channels 
an additional constraint on the range of the central correlator is used ($I_{R_{+}}^{(S=0,T=0)}$=0.1 fm$^4$, cf. Ref.~\cite{Rot.05}).
The details of the determination and parameterization
of the standard correlator are given in Ref. \cite{Rot.05}.
The tensor correlations between two nucleons are generated by a
tangential shift depending
on the spin orientation \cite{Nef.03}. The size and the radial dependence
is given by a tensor correlation
function $\vartheta(r)$ for each of the two $S=1$ channels,
whose parameters are also determined from an energy minimization
in the two-body system \cite{Rot.05}.
The range of the tensor correlation function is restricted through a constraint on the range measure
 $I_{\vartheta}=\int dr\,r^2 \vartheta(r)$.
If one would
use for the description of finite nuclei the long-range tensor
correlator that is optimal for the deuteron, an effective screening due
to other nucleons would emerge through higher-order contributions 
of the cluster expansion~\cite{Nef.03}. In practical
calculations based on two-body approximation, this problem is resolved by restricting the range of the tensor correlation
function, which provides an effective inclusion of the screening effect
without explicit evaluation of  higher terms in the cluster
expansion~\cite{Rot.04}.
Recent studies within the exact no-core shell model~\cite{Rot.05}, show
that $I_{\vartheta}^{(S=1,T=0)}$ = 0.09 fm$^3$ leads to an optimal tensor
correlator for the description of binding energies of $^{3}$H
and $^{4}$He.
In the present work we vary the range of the tensor correlator, $I_{\vartheta}^{(S=1,T=0)}$ = 0.07, 0.08,
and 0.09 fm$^3$, in order to probe its impact on the description of
the global properties of collective excitation phenomena in atomic nuclei.
The contributions of the tensor correlator in $(S,T)=(1,1)$ channel are one
order of magnitude smaller~\cite{Rot.04}, and therefore neglected in
the present study.
\section{Random-phase approximation in the UCOM framework}
Starting from the uncorrelated Hamiltonian for the $A$-body system
consisting of the kinetic energy operator $T$ and a two-body
potential $V$, the formalism of the unitary
correlation operator method is employed 
to generate the correlated Hamiltonian. By combining
the central and tensor correlation operators, the correlated many-body
Hamiltonian in two-body approximation is given by,
\begin{equation}
  \HO_{\UCOM} = {\corr{\TO}}^{[1]} + {\corr{\TO}}^{[2]} + {\corr{\VO}}^{[2]}
  = \TO + \VO_{\UCOM},
\end{equation}
where the one-body contribution comes only from the uncorrelated
kinetic energy ${\corr{\TO}}^{[1]} = \TO$.
Two-body contributions arise from the correlated kinetic
energy ${\widetilde{T}}^{[2]}$ and the correlated potential
${\widetilde{V}}^{[2]}$, which together constitute the correlated
interaction $V_{\UCOM}$~\cite{Rot.05}.  More details about the evaluation
of the two-body matrix elements
for $V_{\UCOM}$ are available in Ref.~\cite{Rot.05}. 
Assuming spherical symmetry, the correlated realistic NN interaction 
is employed to solve the HF equations, i.e. to evaluate the single-particle wave
functions and energies~\cite{RPPH.05}. The UCOM-HF single-particle
spectra are used as a basis for the construction of the $p-h$ configuration space
for the RPA. The RPA equations are derived from 
the equation of motion method using the quasiboson approximation \cite{Row.70},
\begin{equation}
\label{rpaeq}
\left(
\begin{array}{cc}
A^J & B^J \\
B^{^\ast J} & A^{^\ast J}
\end{array}
\right)
\left(
\begin{array}{c}
X^{\nu ,JM} \\
Y^{\nu,JM}
\end{array}
\right) =\omega_{\nu}\left( 
\begin{array}{cc}
1 &  0 \\
0 & -1
\end{array}
\right)
\left( 
\begin{array}{c}
X^{\nu,JM} \\
Y^{\nu,JM}
\end{array}
\right)\; ,
\end{equation}
where the eigenvalues $\omega_{\nu}$ correspond to RPA excitation energies.
The residual particle-hole interaction in $A$ and $B$ matrices includes the correlated
realistic NN interaction $V_{\UCOM}$ in a fully consistent way with the
Hartree-Fock equations. In addition, the multipole transition operators are
consistently transformed by employing the same unitary transformation as for
the Hamiltonian.

However, it turns out that the effect of the UCOM transformation of the transition operators
for monopole and quadrupole modes is negligible~\cite{PPRH.05}.
It is interesting to note,
that the UCOM-RPA
results are in agreement with the study of effective operators in the no-core shell
model within the $2\hbar\Omega$ model space, where the B(E2) values
are very similar for the bare and the effective operator which includes
the two-body contributions~\cite{Ste.05}.
An essential property of the present UCOM-RPA scheme is
that it is fully self-consistent, i.e.
the same correlated realistic NN interaction is used in the HF equations
that determine the single-particle basis, as well as the RPA residual interaction, and the multipole transition operators are
transformed consistently with $\VO_{\UCOM}$.
This means that the same unitary transformation of the realistic NN interaction,
i.e. central and tensor correlation functions with
the same set of parameters are systematically employed in
HF and RPA calculations. The effective NN interaction which determines 
the ground-state properties, also determines
the small amplitude motion around the nuclear ground state.
This property of the present model ensures that RPA amplitudes do not contain
spurious components associated with the center-of-mass translational motion.
Models that are not fully self-consistent necessitate the inclusion of an additional
free parameter in the residual interaction, to adjust a proper separation of 
the spurious state.

One of the interesting questions is to which extent the UCOM-RPA
transition
spectra are sensitive to the range of the tensor correlator employed in the unitary
transformation. We have verified that the multipole strength distributions do
not depend on variations of the central correlator range, around the standard
short-range correlator~\cite{PPRH.05}.

\begin{figure}[th]
\vspace*{20pt}
\centerline{\psfig{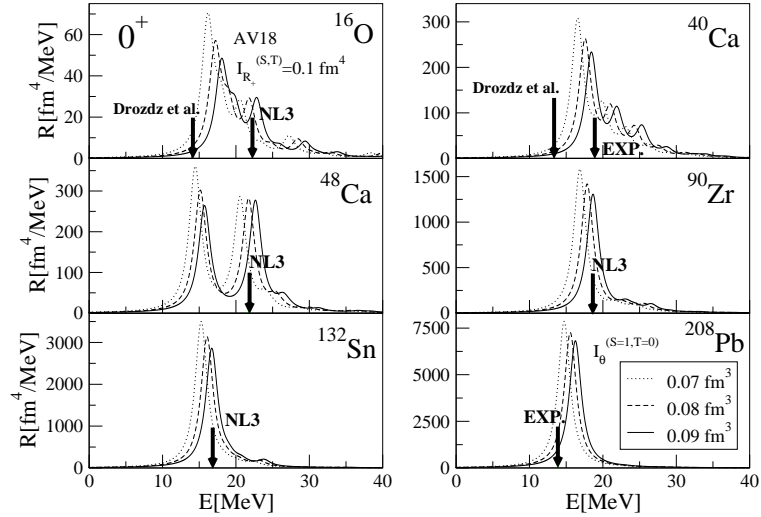}}
\vspace*{8pt}
\caption{The calculated UCOM-RPA strength distribution of ISGMR for the
correlated Argonne V18 interaction, using different restrictions 
on the range of the tensor correlator ($I_{\vartheta}^{(S=1,T=0)}$=0.07,
0.08, and 0.09 fm$^3$). The experimental
data \protect\cite{You.99,Shl.93,Sha.88} and results
from the nonrelativistic (Dro{\. z}d{\. z} et al.) \protect\cite{Dro.90} and
relativistic RPA (NL3) \protect\cite{Paa.03,Ma.01} are denoted by arrows.}
\label{figmono2}
\end{figure}

In Fig.~\ref{figmono2}, we display the UCOM-RPA spectra of isoscalar
giant monopole resonances (ISGMR) for several closed-shell nuclei, using the correlated Argonne
V18 interaction with different constrains on the range of the tensor correlator, $I_{\vartheta}^{(S=1,T=0)}$=0.07, 0.08,
and 0.09 fm$^3$. For comparison, we also denote excitation 
energies from a selection of experimental~\cite{You.99,Shl.93,Sha.88} and
theoretical~\cite{Paa.03,Dro.90,Ma.01} studies.
The agreement with experiment and other theoretical results is rather good for the 
standard correlator set with $I_{\vartheta}^{(S=1,T=0)}$=0.09 fm$^3$. In heavy nuclei, the ISGMR energies are overestimated by $\approx1-3$ MeV.
By varying the range of the tensor correlator around its standard value, the transition strength can be fine-tuned to improve the agreement with the experimental data.

\begin{figure}[th]
\vspace*{22pt}
\centerline{\psfig{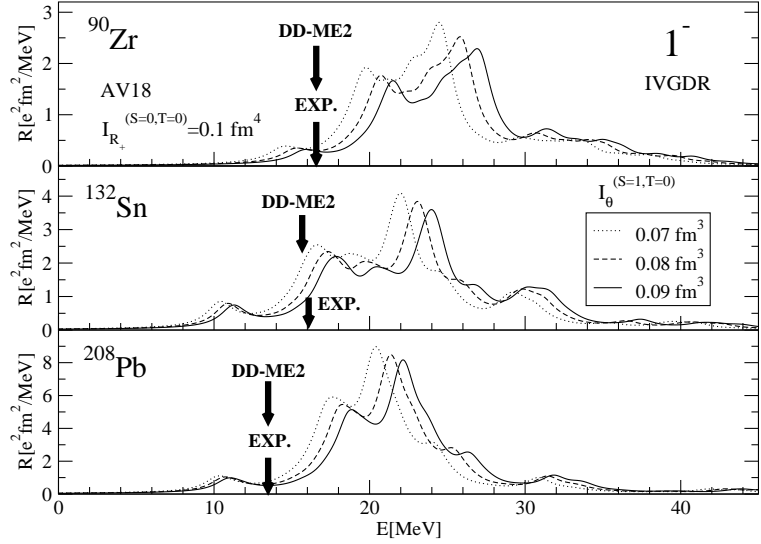}}
\vspace*{8pt}
\caption{The UCOM-RPA strength distributions for the IVGDR in $^{90}$Zr,
$^{132}$Sn, and $^{208}$Pb. The calculations are based on the correlated Argonne V18 interaction, using different constraints on the tensor correlator range ($I_{\vartheta}^{(S=1,T=0)}$= 0.07, 0.08, and 0.09 fm$^3$).
The experimental data~\protect\cite{Adr.05,Ber.75,Poe.89,Rit.93} and
the relativistic RPA (DD-ME2) energies~\protect\cite{NVR.02,LNVR.05} are shown by arrows.}
\label{figdip2}
\end{figure}

Next we employ the UCOM-RPA to describe the isovector giant dipole
resonances (IVGDR)  in $^{90}$Zr, $^{132}$Sn, and $^{208}$Pb (Fig.~\ref{figdip2}). The correlated Argonne V18 interaction is used,
with different constraints on the ranges of the tensor part of the correlator,
$I_{\vartheta}^{(S=1,T=0)}$=0.07, 0.08, and 0.09 fm$^3$.
The calculated dipole response is compared with the 
experimental data~\cite{Adr.05,Ber.75,Poe.89,Rit.93} and with
the theoretical excitation energies from the relativistic
RPA~\cite{NVR.02} based on DD-ME2 interaction~\cite{LNVR.05}.
In all nuclei under consideration, the resulting IVGDR
strength distributions display rather wide resonance-like structures. 
The decrease in the range of the tensor correlator, 
i.e. its constraint $I_{\vartheta}^{(S=1,T=0)}$=0.09 fm$^3$
towards 0.07 fm$^3$, results in lower IVGDR peak energies
by $\approx$2-3 MeV. However, UCOM-RPA overestimates
the IVGDR centroid energies by
$\approx$3-7 MeV. This difference can serve as a
direct measure of the missing correlations
and three-body contributions in the UCOM-RPA scheme.
Inclusion of the three-body interaction
and long-range correlations beyond the simple RPA method,
would probably to a large extent resolve the
presently obtained discrepancies with the other studies. 

\begin{figure}[th]
\vspace*{20pt}
\centerline{\psfig{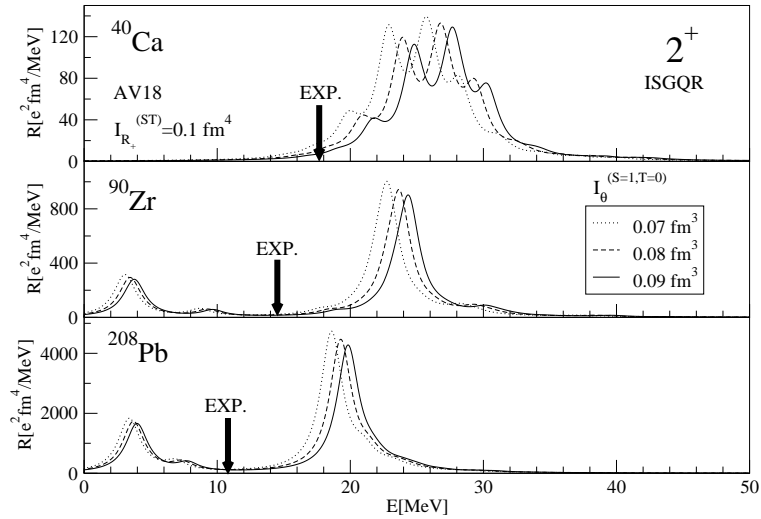}}
\vspace*{8pt}
\caption{The ISGQR strength distributions for
$^{40}$Ca, $^{90}$Zr, and $^{208}$Pb.
The UCOM-RPA model is based on the correlated Argonne V18
interaction with different values of $I_{\vartheta}^{(S=1,T=0)}$=0.07,
0.08, and 0.09 fm$^3$ to constrain the range of the tensor correlator.
The experimental ISGQR excitation energies are denoted
by arrows \protect\cite{Ber.79}.}
\label{figquad2}
\end{figure}

In Fig.~\ref{figquad2} we show the UCOM-RPA isoscalar quadrupole
transition strength distributions for $^{40}$Ca, $^{90}$Zr, and $^{208}$Pb 
(Argonne V18, $I_{\vartheta}^{(S=1,T=0)}$=0.07, 0.08, and 0.09 fm$^3$), in
comparison with experimental data~\cite{Ber.79}.
The residual interaction constructed from the correlated realistic NN
interaction is attractive in the isoscalar channel, generating strongly
collective peaks corresponding to isoscalar giant quadrupole resonance (ISGQR).
In addition, in the case of
$^{90}$Zr, and $^{208}$Pb, the UCOM-RPA model also results with pronounced
low-lying $2^+$ states. However, RPA based on the correlated realistic NN interaction,
without the long-range correlations and three-body contributions,
is not sufficient for a quantitative description of ISGQR excitation 
energy. For the short range tensor correlator ($I_{\vartheta}^{(S=1,T=0)}$=0.07 fm$^3$) the
model still overestimates experimental values by $\approx$ 8 MeV. By decreasing
the range of the tensor correlator, the quadrupole response is systematically
shifted towards lower energies.  The quadrupole response is rather
sensitive to the range of the tensor correlator. For $^{40}$Ca and the ranges of tensor
correlator determined by $I_{\vartheta}^{(S=1,T=0)}$=0.07, 0.08, and 0.09 fm$^3$,
the ISGQR centroid energies read 25.1, 26.2, and 27.1 MeV, respectively.
In the cases of heavier nuclei, these differences are smaller, e.g. for $^{208}$Pb,
the centroid energy lowers by 1.2 MeV when going from the correlator
with $I_{\vartheta}^{(S=1,T=0)}$=0.09 fm$^3$ towards $I_{\vartheta}^{(S=1,T=0)}$=0.07 fm$^3$.

The agreement achieved between the calculated and experimental 
properties of the ISGMR indicates that the correlated NN interaction
corresponds to realistic values of the nuclear matter (NM) incompressibility. 
It has been demonstrated in the past that, within 
relativistic and non-relativistic RPA, 
the energies of the dipole and quadrupole resonances, on one hand, 
and the value of the effective mass corresponding to the effective 
interaction used, on the other, are correlated~\cite{Hui.89,Rei.99}.
In particular, the relativistic RPA without density-dependent 
interaction terms, based on the ground state with a small effective
mass and relatively high compression modulus, 
resulted in systematically overestimated energies of 
giant resonances~\cite{Hui.89}.
The discrepancies between UCOM-RPA calculations and experimental
data for multipole giant resonances, as well as the low density
of single-nucleon UCOM-HF states, suggest that the respective
effective mass is too small. Tensor correlations with shorter range
increase the single-particle level density and result in a systematic
shift of the giant resonances towards lower energies, improving the
agreement with experimental data. 
However, the variations of the ranges of correlation functions can
serve only as a tool for 
fine tuning of the excitation spectra and they can not supplement
the effects of the missing long-range correlations and
three-body contributions.

\section{Conclusions}
In the present study, the fully self-consistent RPA is formulated
in the single-nucleon Hartree-Fock basis by using correlated 
realistic NN interactions. The short-range central and tensor 
correlations induced by the NN potential are explicitly treated
within the UCOM
framework.
It is shown that the correlated NN interactions are successful in
generating collective excitation modes, but for an
accurate description of experimental data on excitation energies and transition
strengths, one has to account for the contributions missing in the present treatment. These are (i) long-range correlations beyond simple
RPA, which can be included within a RPA scheme built on the correlated ground state or by including complex configurations within Second-RPA, and (ii) induced and genuine three-body interactions, which one could try to model by a simple effective three-body force.  

\section*{Acknowledgments}
This work has been supported by
the Deutsche Forschungsgemeinschaft (DFG) under contract SFB 634. We thank the Institute for Nuclear Theory at the University of
Washington for its hospitality and the Department of Energy for partial support 
during the completion of this work.
\end{document}